\documentclass[aps,prb,twocolumn,groupedaddress,eqsecnum,showpacs]{revtex4} 
\usepackage{amsfonts}
\usepackage{dcolumn}
\usepackage{graphicx,amssymb,amsmath}
\usepackage{bm}
\usepackage{natbib}
\usepackage{epstopdf}
\begin{document}

\title{Quasiparticle excitations in Bose-Fermi mixtures}
\author{Shimul Akhanjee}
\email[]{shimul@riken.jp}

\affiliation{Condensed Matter Theory Laboratory, RIKEN, Wako, Saitama, 351-0198, Japan}


\date{\today}

\begin{abstract}
We analyze the excitation spectrum of a three-dimensional(3D) Bose-Fermi mixture with tunable resonant interaction parameters and high hyperfine spin multiplets. We focus on a 3-particle vertex describing fermionic and bosonic atoms which can scatter to create fermionic molecules or disassociate. For a single molecular level, in analogy to the single magnetic impurity problem we argue that the low lying excitations of the mean-field theory are described by the Fermi liquid picture with a quasiparticle weight and charge which is justified by a $1/N_\psi$ expansion, expected to be exact in the limit of infinite degeneracy (or very high fermionic spin) $N_\psi \to \infty$. Our emphasis is placed on the novel conditions for chemical equilibrium and how many-body chemical reactions renormalize the bosonic chemical potential, modifying condensation and superfluid-insulator transitions. 
\end{abstract}


\maketitle

\section{introduction}
\begin{figure}
\centerline{\includegraphics[width=2.7in]{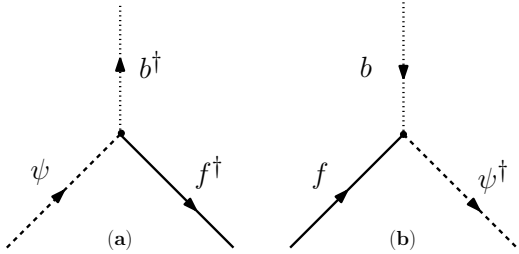}}
\caption{Vertices associated with the interaction term $H_{bf}^{3p}$ (a) Disassociation of a fermionic molecule into bosonic and fermionic atoms (b) Molecular formation through tunable resonant scattering.  }
\label{fig:vert}	
\end{figure}
Recent experimental studies of ultra cold atomic gases cooled to quantum degeneracy and trapped in optical lattice potentials have demonstrated precise control over two-body interactions\cite{blochrmp08}. The Feshbach resonance, which can be exploited by ramping a magnetic field $\Delta B$ near the resonance $B_0$, can tune the scattering length $a_0$ over a wide range of values, generating both repulsive and attractive interactions (of strength $g_{\alpha,\beta}$ ) between pairs of atoms $\alpha$ and $\beta$. The interaction coupling has the following dependence on key quantities, 
\begin{equation}
g_{\alpha \beta }  \approx \sqrt {\frac{{2\pi a_{bg} \Delta B\delta \sigma _{\alpha \beta } }}{{m_\psi  }}} 
\label{eq:gdef}
\end{equation}
where $\delta \sigma _{\alpha \beta }$ is a field dependent quantity that depends on the mismatch of the magnetic moments and $a_{bg}$ is the background scattering length. 

Furthermore, optical lattice systems have provided a remarkably ideal testing ground for new strongly correlated phenomena. 
In particular, Bose-Fermi mixtures possess a unique physical realization beyond the scope of traditional solid state materials. Several experimental groups have successfully synthesized mixtures of $^6$Li Fermions with both $^7$Li Bosons\cite{truscott,schreck} and $^{23}$Na Bosons\cite{stan}, and more recently  $^{40}$K Fermions with $^{87}$Rb Bosons\cite{inouye}. The electrically polar nature of KRb and other heteronuclear diatomic molecules allow for novel quantum states driven by electric dipole-dipole interactions\cite{hudson}, however for simplicity such effects will not be discussed here. Most recent experiments have reported interspecies Feshbach resonances in mixtures
of Rb and Cs atoms\cite{pilchPRA2009}. 

Earlier theoretical studies have focused on density-density type interspecies interactions. Interesting phenomena utilizing various mean-phase studies\cite{adhikariPRA04} has found the precise limits of phase separation\cite{molmerPRL98,amoruso98,buchlerPRA04}. Certain one-dimensional approaches\cite{kkdasPRL03,cazalillaPRL03} have predicted novel 1D ordered phases where strong coupling theories\cite{wangPRL2006} seem to break down. More recent numerical work by Lewenstein \emph{et al.}\cite{lewensteinPRL2004} identified several pairing instabilities with a rich phase diagram consisting of various paired phases between the bosons and fermions, with order ranging from charge density wave to superconductivity. Later studies introduced a 3-particle interaction which explicitly allows for the creation of a boundstate molecular particle\cite{kokPRA2002,pethickPRL2004,bortJPB2006,bortPRA08}, revealing Bose-Einstein condensate depletion and other manifestations of phase separation\cite{powellPRB05,bortPRA08}. In this article we shall focus on the consequences of this type of interaction.

Unlike simple electrons, atoms and molecules have internal degrees of freedom that build composite particles with total spin given by $F=I+S$, where $I$ is the nuclear spin and $S$ is the electron spin. In the case of fermions, the system can exhibit high degeneracy due to hyperfine spin multiplets such as the recently observed $^{173}$Yb Fermi gas, with constituents having spin of $5/2$\cite{fukuhara}. Therefore, here we consider such a system, consisting of a degenerate Bose-Fermi mixture containing high spin fermions of degeneracy $N_\psi = 2F+1$, which can combine with bosonic atoms via resonant scattering. We derive a Fermi-liquid picture of the low-lying excitations of a single molecular level, having an effective quasiparticle weight and broadening width, taking in account non-Gaussian fluctuations at the order $1/N_\psi$. The analysis presented here fundamentally unifies the condensed matter physics with atomic physics and chemistry, given that the molecular levels which are hybridized through a many-body chemical reaction can be described through the quasiparticle concept. Thus, if the chemical equilibrium conditions can be violated, the components of the mixture will be driven by a steady reaction rate that will deplete one or more atomic or molecular species either by net molecular formation or disassociation.

\section{The model}
Here we study a model of a mixture of spin-polarized fermionic atoms and bosonic atoms, both of which can be treated as spinless particles. The canonical annihilation and creation operators are defined for the Bosons - ${\hat b},{\hat b}^\dag$ and Fermions - ${\hat f},{\hat f}^\dag$, combined to form fermionic diatomic molecules ${\hat \psi},{\hat \psi}^\dag$. The most general Hamiltonian can be decomposed into the following parts:
\begin{equation}
\mathcal{H} = \mathcal{H}_{0} + \mathcal{H}_{bb} + \mathcal{H}_{bf}^{2p} + \mathcal{H}_{bf}^{3p}
\label{eq:hamilt1} 
\end{equation}
where, 
\begin{eqnarray}
{\mathcal{H}_{0}} &=& \sum\limits_{\vec k} {\varepsilon _{\vec k}^f\hat f_{\vec k}^\dag {{\hat f}_{\vec k}}}  + \sum\limits_{\vec k} {\varepsilon _{\vec k}^b\hat b_{\vec k}^\dag {{\hat b}_{\vec k}}}  + \sum\limits_{\vec k} {\varepsilon _{\vec k}^\psi \hat \psi _{\vec k}^\dag {{\hat \psi }_{\vec k}}} \\ 
 {\mathcal{H}_{bb}} &=& {g_{bb}}\sum\limits_{\vec k,\vec k',\vec q} {\hat b_{\vec k' + \vec q}^\dag \hat b_{\vec k - \vec q}^\dag {{\hat b}_{\vec k}}{{\hat b}_{\vec k'}}}  \\ 
\mathcal{H}_{bf}^{2p}&=& g_{bf}^{2p}\sum\limits_{\vec k,\vec k',\vec q} {\hat f_{\vec k' + \vec q}^\dag \hat b_{\vec k - \vec q}^\dag {{\hat f}_{\vec k}}{{\hat b}_{\vec k'}}}  \\ 
\mathcal{H}_{bf}^{3p}&=& g_{bf}^{3p}\sum\limits_{\vec k,\vec k',\vec q} {\hat \psi _{\vec k + \vec k'}^\dag {{\hat f}_{\vec k}}{{\hat b}_{\vec k'}} + \hat b_{\vec k'}^\dag \hat f_{\vec k}^\dag {{\hat \psi }_{\vec k + \vec k'}}}  
\end{eqnarray}
In a ultra cold atomic systems the particles are usually confined by a superposition of harmonic trapping potentials. The resulting momentum basis $\vec k$ is defined in the Bloch representation, commensurate with the periodicity of the optical lattice. In this basis the resulting particle dispersions are given by 
\begin{equation}
\varepsilon _{\vec k}^{\nu  = b,f,\psi } = \frac{{{\hbar ^2}{{\vec k}^2}}}{{2{m_\nu }}} - {\mu _\nu }
\end{equation}
where the respective chemical potentials $\mu_\nu$ should obey the sum rule:
\begin{equation}
\sum\nolimits_{i = reac\tan ts} {{\mu _i}}  = \sum\nolimits_{j = products} {{\mu _j}} 
\end{equation}
for chemical equilibrium. 

Since the $\mu_\nu$'s are fixed, the particle number can fluctuate as if the system is coupled to external particle resevoirs, and the analysis is performed in the grand canonical ensemble for simplicity although more realistic models should hold the particle number fixed. Consequently, an important point to consider is that the special form of $ \mathcal{H}^{3p}$ breaks $U(1)$ gauge invariance under the tranformations $\hat b \to \hat b e^{i\theta}$,  $\hat f \to \hat f e^{i\theta}$,  $\hat \psi \to \hat \psi e^{i\theta}$. A related issue follows from the commutator $\left[ {{\mathcal{H}_{bf}^{3p}},\hat N} \right] \ne 0$ with the number operator defined by $\hat N = \hat b^\dag {\hat b} + \hat f^\dag {\hat f} + \hat \psi ^\dag {\hat \psi }$.  Evidently this implies that particle number is not conserved, which is not unexpected as the formation and disassociation of molecular bound-states should not conserve particle number.

The fermionic s-wave scattering is usually suppressed due to the Pauli exclusion principle and can be omitted from the outset. Furthermore, we take the regime of net attraction $g_{bf}^{2p} << g_{bf}^{3p}$ such that $H_{bf}^{2p}$ can be neglected.  
In order to obtain controlled and qualitatively simple analytical results when integrating over momenta, we shall invoke a wide flat-band approximation where the two species of Fermions occupy separate kinetic bands with separate Fermi surfaces, and the effective masses are large enough such that the dispersion relations are slow varying in momentum space near the bottom of the bands:  $\varepsilon _{\vec k}^f \approx \varepsilon _{0}^f -\mu_f$, $\varepsilon _{\vec k}^\psi \approx \varepsilon _{0}^\psi -\mu_\psi$. Although the fermionic atoms and molecules occupy decoupled Fermi surfaces the hybridization effects of the molecules with the atoms will manifest through the broadening of the molecular Fermi energy. 

The Bosons are sufficiently condensed into a BEC phase such that the total number of Bosons is ${N_b ^0} \simeq {L^3}\hat b_0^\dag \hat b_0^\dag $, for a restricted volume $L^3$.  It follows that the two-body and three-particle Bose-Fermi couplings $g_{bf}^{2p}$ and $g_{bf}^{3p}$ can be tuned by a magnetic field via Eq.(\ref{eq:gdef}), thus driving the formation of molecular bound states or renormalizing the kinetic bands. The vertices associated with the $g_{bf}^{3p}$ type of interaction are shown in Fig.\ref{fig:vert}.
\begin{figure}
\centerline{\includegraphics[width=3.0in]{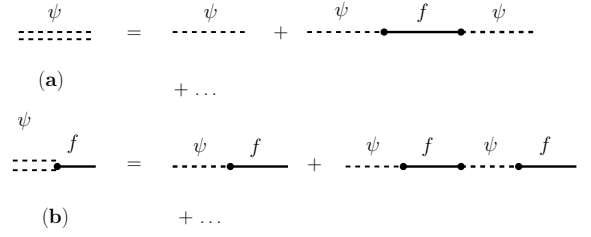}}
\caption{(a) The hybridized, one particle Green's function, $G_{\psi} ^{MF} (i\omega_n)$ expanded in powers of the bosonic mean field. (b) The terms leading to the mixed amplitude $G_{\psi,f} ^{MF} (i\omega_n)$.  }
\label{fig:mf}	
\end{figure}
Term $H_{bb}$ describes Boson-Boson interactions, which affect the condensation fraction $N_b ^0 /N_b$ at low temperatures. A sharp transition into the BEC phase occurs when the chemical potential reaches a critical threshold at which $\mu_b ^c = E_{b} ^0$, where $E_{b} ^0$ is the lowest energy that the Bosons can occupy. This transition can be understood as the breaking of the $U(1)$ gauge symmetry associated with the change of phase in the many-body wave function. For free Bosons in a box, $\mu_b ^c = 0$, however in a harmonic trap $\mu _b^c = (\hbar /2)({\omega _1} + {\omega _2} + {\omega _3})$, where $\omega _i$ is the trapping frequency along the $i$th direction. Hence, in physically realizable ultra cold atomic systems, the condensate fraction $N_b ^0 /N_b$ and $\mu_b ^c$ strongly depend on the details of the confining potentials in addition to the 2 and 3 body interaction effects. If the condensate fraction is large enough, the Bogoliubov approximation\cite{abrikosov} can be employed, resulting in a renormalized Boson mass $m_b$ and dispersion relation which can be redefined in the original Hamiltonian. 

\section{Mean field theory}
\subsection{Derivation of the quasiparticle Hamiltonian} 
Here we investigate a special case of the system described by the Hamiltonian (\ref{eq:hamilt1}) and restrict our analysis to the degenerate molecules at the Fermi surface that hybridize with the same number of flat-band atomic Fermions. We can express the partition function as a functional integral,
\begin{equation}
\mathcal{Z} = \int {\mathcal{D}\hat b\mathcal{D}{{\hat b}^\dag }\mathcal{D}\hat f\mathcal{D}{{\hat f}^\dag }\mathcal{D}\hat \psi \mathcal{D}{{\hat \psi }^\dag }\exp \left( { - \int_0^\beta  {\mathcal{L}(\tau )d\tau } } \right)} 
\end{equation}
with the Lagrangian given by
\begin{equation}
\begin{aligned}
 \mathcal{L}(\tau ) &= \\
&=\hat b_0^\dag \left( {\frac{\partial }{{\partial \tau }} + \varepsilon _0^b} \right){{\hat b}_0} + \sum\limits_{k,m} {\hat f_{km}^\dag \left( {\frac{\partial }{{\partial \tau }} + \varepsilon _k^f} \right){{\hat f}_{km}}  }  \\ 
 &+\sum\limits_m^{{N_\psi }} {\hat \psi _m^\dag \left( {\frac{\partial }{{\partial \tau }} + \varepsilon _F^\psi } \right)} {{\hat \psi }_m}\\ &+ V\sum\limits_{km} {\hat \psi _m^\dag {{\hat f}_{km}}\hat b 
+ {{\hat b}^\dag }\hat f_{km}^\dag {{\hat \psi }_m}}  + {\mathcal{H}_{bb}}  
 \end{aligned}
 \label{eq:mflagrange}
\end{equation}
where the constant $V$ is proportional to $g_{bf}^{3p}$ and the degeneracy index $m$ is summed to $N_\psi$. Note that the degeneracy index which is attached to the operator ${\hat f}_{km}$ reflects stoichiometry, in that the number of fermions that participates in the hybridization is fixed by the molecular level degeneracy. The resulting model described by Eq.(\ref{eq:mflagrange}) has a form that at first glance mirrors the slave boson mean-field theory of the infinite $U$ Anderson model, for which the mean-field theory and $1/N$ corrections have been thoroughly studied. However the operator constraint which projects out the Bosons\cite{hewsonbook} is replaced by the bosonic repulsion and chemical potential terms. Nevertheless, the two models exhibit distinct phenomena which will be clarified in a later section.

Consider the effect of integrating out either both or one of the $\hat f$ and $\hat \psi$ fields. Since the $\mathcal{H}_{bf}^{3p}$ is linear all fields, we can make use of the Hubbard-Stratonovich transformation. A proper choice of the integration depends on dominant low energy scale of each particle species. Ref.\cite{bortPRA08} for instance, integrates out only the molecular field $\psi$ as follows, 
\begin{equation}
{\mathcal{Z}_{eff}^{2p}} = \mathcal{Z}_0^{\psi {\mathop{\rm int}} }\mathcal{Z}_0^{b,f}\int {\mathcal{D}\hat b\mathcal{D}{{\hat b}^\dag }\mathcal{D}\hat f\mathcal{D}{{\hat f}^\dag }{{e^{ - S_{eff}^{2p} - \int_0^\beta  {{\mathcal{H}_{bb}}(\tau )d\tau } }}}}
\end{equation}
where $\mathcal{Z}_0^{\psi {\mathop{\rm int}} }$ is an irrelevant constant and $\mathcal{Z}_0^{b,f}$ is the partition function of the non-interacting $f$,$b$ systems. The effective action becomes
\begin{equation}
S_{eff}^{2p} = \int_0^\beta  {{V^2}\sum\limits_{km} {{{\hat f}_{km}}\hat b{{\left( {\frac{\partial }{{\partial \tau }} + \varepsilon _k^\psi } \right)}^{ - 1}}{{\hat b}^\dag }\hat f_{km}^\dag } d\tau } 
\end{equation}
Thus, the resulting attractive 2-body interaction simply offsets the repulsive two-body coupling constant $g_{bf}^{2p}$. This binding energy reaches a threshold at which the scattering amplitude vanishes. Beyond this zero energy regime are 3-particle processes shown in Fig. \ref{fig:vert} which is the central focus of this article. 

On the other hand, if the $f$ fields are integrated out followed by the $\psi$ fields, then the partition function becomes
\begin{equation}
\mathcal{Z}_{eff}^b = \mathcal{Z}_0^f\int {\mathcal{D}\hat b\mathcal{D}{{\hat b}^\dag }\exp \left( { - S_{eff}^b} \right)} 
\end{equation}
with 
\begin{equation}
\begin{array}{l}
 S_{eff}^b =  - N_\psi Tr\ln \left[ {\frac{\partial }{{\partial \tau }} + \varepsilon _F^\psi  - {V^2}{{\hat b}_0}\sum\limits_{k,m} {{{\left( {\frac{\partial }{{\partial \tau }} + \varepsilon _k^f} \right)}^{ - 1}}} \hat b_0^\dag } \right] \\ 
  + \int_0^\beta  {d\tau \hat b_0^\dag \left( {\frac{\partial }{{\partial \tau }} + \varepsilon _0^b} \right){{\hat b}_0}}  + {\mathcal{H}_{bb}} \\ 
 \end{array}
\end{equation}
where $\mathcal{Z}_0^f$ is the non-interacting partition function of the fermionic atoms. By approximating this functional integral by the maximum value of its integrand for $\tau$ independent bosonic fields, we arrive at our version of the mean-field approximation, where the $\hat b$'s of the low energy ground state will be replaced by their expectation values: $\hat b_0= \left\langle {\hat b_0} \right\rangle  = r$ and ${{\hat b}_0 ^\dag } = \left\langle {{{\hat b_0}^\dag }} \right\rangle  = r$, with $r$ defined as a variational mean-field parameter that minimizes the free energy $\mathcal{F} =-T\ln \mathcal{Z}$. After applying this tranformation the following mean field Hamiltonian becomes,
\begin{equation}
\mathcal{H}_{mf} = \mathcal{H}_{qp} - \mu_b r^2 + g_{bb} r^4
\end{equation}
with the quasiparticle Hamiltonian given by the expression,
\begin{equation}
\begin{aligned}
 &{\mathcal{H}_{qp}} = \\
&\sum\limits_k {\varepsilon _{\vec k}^f\hat f_{\vec k}^\dag {{\hat f}_{\vec k}}}  + \sum\limits_m {\varepsilon _{F}^\psi \hat \psi _{m}^\dag {{\hat \psi }_{m}}}
  + \sum\limits_{m,\vec p} {{{\tilde V}}\hat \psi _{m}^\dag {{\hat f}_{m,\vec p }} + \tilde V^*\hat f_{m,\vec p}^\dag {{\hat \psi }_{m}}}  
\end{aligned}
 \label{eq:mfham}
\end{equation}
with an effective molecule-Fermion hybridization coupling ${\tilde V = Vr}$. 
\subsection{Quasiparticle Thermodynamics}
The quasiparticle free energy $\mathcal{F}_{qp}$ can be determined from the partition function,
\begin{equation}
\begin{aligned}
 {\mathcal{F}_{qp}} &=  - \frac{1}{\beta }\ln {\mathcal{Z}_{qp}} \\ 
  &=  - \frac{{{N_\psi }}}{\beta }\int_{ - \infty }^\infty  {\ln \left( {1 + {e^{ - \beta \omega }}} \right){\rho _{qp}}(\omega )d\omega }  \\ 
 \end{aligned}
 \label{eq:free1}
\end{equation}
where the quasiparticle density of states is given by
\begin{equation}
{\rho _{qp}}(\omega ) = \sum\nolimits_n {\delta (\omega  - \varepsilon _n^{qp})}  = \frac{1}{\pi }{\mathop{\rm Im}\nolimits} \left[ {\frac{\partial }{{\partial \omega }}\ln {G_{qp}}({\omega ^ + })} \right]
\label{eq:qpdens}
\end{equation}
and the single quasiparticle Green's function $G_{qp}(i{\omega _n})$ can be determined by solving the coupled equations for a broadened energy level that often appears in the analysis of local magnetic moments in metals,
\begin{equation}
{G_{qp}}{(i{\omega _n})^{ - 1}} = {\left( {i{\omega _n} - \varepsilon _F^\psi  - {{\left| {\tilde V} \right|}^2}\sum\limits_k {\left( {i{\omega _n} - \varepsilon _k^f} \right)} } \right)^{ - 1}}
\label{eq:mfgreen} 
\end{equation}

After substituting Eq.(\ref{eq:qpdens}) via Eq.(\ref{eq:mfgreen}) into Eq.(\ref{eq:free1}) and integrating by parts we have, 
\begin{equation}
\begin{array}{l}
 {F_{MF}} =  - \frac{{{N_\psi }}}{\pi }\sum\limits_n {\ln \left( { - i{\omega _n} + \varepsilon _{F}^\psi  + \sum\limits_{\vec k} {\frac{{{{\left| {{{\tilde V}}} \right|}^2}}}{{\left( {i{\omega _n} - \varepsilon _{\vec k}^f} \right)}}} } \right)} \\
-{\mu_b}{r^2} + {g_{bb}}{r^4}\\ 
  =  - \frac{{{N_\psi }}}{\pi }\int\limits_{ - \infty }^\infty  {\bar f(\omega ){\mathop{\rm Im}\nolimits} [\ln {G_\psi ^{MF} }(i{\omega _n})]d\omega } -{\mu_b}{r^2} + {g_{bb}}{r^4} \\ 
 \end{array}
\end{equation}
where the quasiparticle distribution $\bar f(\omega)$ is given by,
\begin{equation}
\bar f(\omega ) =  \frac{1}{{{e^{\beta \omega  }} + 1}}
\end{equation}
and the low temperature behavior follows from a Sommerfeld expansion of the $\bar f(\omega )$ distribution,
\begin{equation}
\begin{array}{l}
 \int\nolimits_{ - \infty }^\infty  {\bar f(\omega )H(\omega )d\omega }  \simeq \\
 \int\nolimits_{ - \infty }^{{\mu _{qp}}} {H(\omega )d\omega }   
  + \frac{{{\pi ^2}}}{6}{T^2}{{\left. {H'(\omega )} \right|}_{\omega  = {\mu _{qp}}}}  +\mathcal{O}(T^4)
 \end{array}
\label{eq:somfeld}
\end{equation}
where $H(\omega)$ is a function that vanishes at as $\omega \to 0$. 
Next we utilize the wide band limit approximation, where the integration is taken over a width $2W$ with a constant density of states $\rho_0$, and the summation reduces to $\sum\nolimits_{\vec k} {{{\left| {{{\tilde V}_{\vec k}}} \right|}^2}/(i{\omega _n} - \varepsilon _{\vec k}^f)}  =  - i\Delta {\mathop{\rm sgn}} ({\omega _n})$, yielding

\begin{equation}
\begin{aligned}
&{F_{MF}} = \\ &- \frac{{{N_\psi }}}{\pi }\int\nolimits_{ - W}^W {\bar f(\omega ){{\tan }^{ - 1}}\left( {\frac{r^2\Delta }{{\varepsilon _{F}^\psi  - \omega }}} \right)d\omega } -{\mu_b}{r^2} + {g_{bb}}{r^4}
\end{aligned}
\label{eq:mfree1}
\end{equation}
and the ground state energy ($T=0$) evaluates to,
\begin{equation}
\begin{array}{l}
 {E_{gs}} = -\frac{{N_\psi \tilde \Delta }}{\pi } + \frac{{N_\psi \varepsilon _F^\psi }}{\pi }{\tan ^{ - 1}}\left( {\frac{{\tilde \Delta }}{{\varepsilon _F^\psi }}} \right) \\ 
  + \left( {\frac{{N_\psi \tilde \Delta }}{{2\pi }}} \right)\ln \left( {\frac{{{{(\varepsilon _F^\psi )}^2} + {{\tilde \Delta }^2}}}{{{W^2}}}} \right) - {\mu_b}\frac{{\tilde \Delta }}{\Delta } + {g_{bb}}{\left( {\frac{{\tilde \Delta }}{\Delta }} \right)^2} \\ 
 \end{array}
\end{equation}
with $\tilde \Delta = r^2 \Delta$ determined through a minimization that solves the equation,
\begin{equation}
\frac{{\partial {E_{gs}}}}{{\partial \tilde \Delta}} = \frac{N_\psi}{{2\pi }}\ln \left( {\frac{{{{(\varepsilon _F^\psi )}^2} + {{\tilde \Delta }^2}}}{{{W^2}}}} \right) - \frac{\mu_b}{\Delta } + \frac{{2{g_{bb}}\tilde \Delta }}{{{\Delta ^2}}} = 0
\label{eq:min}
\end{equation}

Next, we invoke the Friedel sum rule, which relates the quasiparticle occupation to the self-energy corrections and broadening factor,\cite{hewsonbook}, 
\begin{equation}
{n_{\varepsilon _F^\psi }} = (N/\pi ){\tan ^{ - 1}}(\tilde \Delta /\varepsilon _F^\psi )
\end{equation}
Consequently, in order to maintain a finite density of levels as $N_\psi \to \infty$, then $\tilde \Delta \to 0$. Taking this limit in Eq.(\ref{eq:min}) leads to the following zero temperature shift of the bosonic chemical potential,
\begin{equation}
\delta {\mu _b}(T=0) = \frac{{N_\psi \Delta }}{\pi }\ln \left( {\frac{{k_B  T_F ^\psi }}{W}} \right)
\label{eq:mffinal}
\end{equation}
with the thermal energy scale of the molecules defined by $\varepsilon _F^\psi \equiv k_B T_F ^\psi$ which is valid provided that $\varepsilon _F^\psi \ll W$. Evidently, the BEC will be depleted if this correction shifts the bosonic chemical potential above its lowest available energy. Moreover, the low temperature behavior can be obtained by applying the expansion of Eq.(\ref{eq:somfeld}) in the wide band limit to yield,
\begin{equation}
\delta {\mu _b}(T) \simeq \delta {\mu _b}(0) + \frac{{\pi N_\psi\Delta }}{6}{\left( {\frac{T}{{{T_F ^\psi }}}} \right)^2} + \mathcal{O}({T^4})
\end{equation}
Thus, these corrections determine the large $N_\psi$, low $T$ behavior as compared to the characteristic temperature $T_F ^\psi$. However, for a proper sense of the behavior near $T_F ^\psi$, it is necessary to go beyond the mean field theory.

\subsection{Quasiparticle weight and Fermi liquid theory}
In order to formally establish a local Fermi-liquid picture for a broadened molecular level, one must determine the quasiparticle weight and establish a 1 to 1 correspondence between the quasiparticles and original molecular levels. The propagator of the broadened molecular Fermi level is related to the quasiparticle propagator by the relation,
\begin{equation}
\begin{array}{l}
 {G_{{\varepsilon _\psi }}}(\omega ) = {\left| {\left\langle b \right\rangle } \right|^2}{G_{qp}}(\omega ) \\ 
  = {\left| {\left\langle b \right\rangle } \right|^2}{\left( {i{\omega _n} - \varepsilon _F^\psi  - {{\left| {\tilde V} \right|}^2}\sum\limits_k {\left( {i{\omega _n} - \varepsilon _k^f} \right)} } \right)^{ - 1}} \\ 
 \end{array}
\end{equation}
In the context of the theory of the Fermi liquid, ${\left| {\left\langle b \right\rangle } \right|^2} = r^2$ plays the role of the wave function renormalization factor or quasiparticle weight $z$. Furthermore, $z$ can be related to several conservation laws that arise from constraints on the volume of the Fermi surface. Recent studies by Powell \emph{et al.} have dervied such laws given for $\mathcal{H}_{3p}$ by applying Luttinger's theorem which states that the volume of the respective Fermi surfaces is conserved even in the presence of interactions\cite{powellPRB05}. These conservation laws take the form,
\begin{equation}
\begin{array}{l}
 z = {\left| {\left\langle b \right\rangle } \right|^2} = {N_b} - {\left| {\left\langle \psi  \right\rangle } \right|^2} \\ 
 {\left| {\left\langle f \right\rangle } \right|^2} = {N_f} - {\left| {\left\langle \psi  \right\rangle } \right|^2} \\ 
 \end{array}
\end{equation}

Additionally, by applying the relation above one can then generate conserved charges. Formally, with any conservation law there is an associated quantum number $Q$ such that quasi particle charges can be defined. However, $Q$ should vanish if the particle-hole symmetry of the system is strictly upheld. More importantly, the validity of these conservation laws can be tested via the measurement of $z$ and $Q$, which should be directly accessible from spectral line broadening measurements and laser absorption techniques that can capture velocity and density distribution profiles to a high degree of precision\cite{blochrmp08}.

\subsection{Diagrammatic approach}
\begin{figure}
\centerline{\includegraphics[width=2.5in]{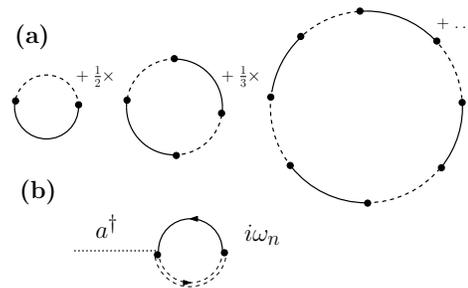}}
\caption{Equivalent contributions to the mean-field free energy of order $N_\psi$. (a) Ring diagram summation and (b) Hartree approximation. }
\label{fig:ring}	
\end{figure}

An alternative approach, which will be useful in considering non-Gaussian fluctuations is to derive the preceding results diagrammatically from the standard Green's function technique\cite{abrikosov}, involving an expansion in powers of the hybridization $\tilde V$.   
The resulting thermal, one-particle propagator with Lorentzian broadening can be computed from the series in Fig.\ref{fig:mf}(a). Additionally, another propagator of interest describes the mixed molecule-Fermion amplitude $\left\langle {\mathcal{T}{{\hat f}_{m,\vec k}}\hat \psi _{m}^\dag } \right\rangle =G_{\psi,f} ^{MF} (i\omega_n)$, which can be determined by Fig. \ref{fig:mf}(b) as,
\begin{equation}
\begin{aligned}
 {G_{\psi ,f}^{MF}}{(i{\omega _n})^{ - 1}} &= \\
 =\left( {i{\omega _n} - \varepsilon _{\vec k}^f} \right)& \left( {i{\omega _n} - \varepsilon _F^\psi  - \sum\limits_k {\frac{{{{\left| {{{\tilde V}_{\vec k}}} \right|}^2}}}{{\left( {i{\omega _n} - \varepsilon _{\vec k}^f} \right)}}} } \right) \\ 
\end{aligned}
\label{eq:mixed}
\end{equation}
Next we can proceed to determine the change in mean-field free energy of order $N_\psi$ due to $\mathcal{H}_{bf}^{3p}$. The mean-field equations follow from a perturbation series in the usual way involving Feynman diagrams. The first approach follows from the linked cluster expansion where the mean-field free energy is computed from the sum of the ring diagrams as shown in Fig.\ref{fig:ring}(a). The second approach follows from the Hartree tadpole diagram as shown in Fig. \ref{fig:ring}(b), which is made to cancel the external field terms, reducing to the desired mean-field equation. In both cases we recover Eq.(\ref{eq:min}).

\section{Beyond mean-field theory}
Previously in the mean field solution, $\hat b$ and $\hat b^\dag$ were replaced by their expectation values. Beyond mean field theory we make use of the transformations,
\begin{equation}
\hat b = \hat a + c_0, \,\,\,\,{{\hat b}^\dag }  = \hat a^\dag + c_0
\end{equation}
where $\hat a$ and ${\hat a}^\dag$ are static operators that obey the Bose commutation rules and $c_o$ is a constant. After making these replacements, the Hamiltonian separates into two parts consisting of the mean-field part given by Eq.(\ref{eq:mfham}) and a fluctuating part containing the rest of the terms.
 
\subsection{Hartree approximation}

The Hartree approximation can be obtained by analyzing the tadpole diagram of Fig.\ref{fig:ring}(b), which contains a closed loop of the mixed amplitude $G_{\psi,f} ^{MF} (i\omega_n)$, represented by Eq.(\ref{eq:mixed}) and a static $\hat a^\dag$ interaction line. Here the mean-field equation can be derived by choosing the external field $r$ such that it cancels the the tadpole diagram. This condition is satisfied by adjusting the terms that are linear in the $\hat a$ and ${\hat a}^\dag$ fields to zero as follows:
\begin{equation}
{\mu _b}r + 2{g_{bb}}{r^3} + \sum\limits_{n,\vec k} {{{\tilde V}_{\vec k} }{G_{\psi ,f} ^{MF}}(i{\omega _n})}  = 0
\end{equation}
After carrying out the frequency summation and $k$ integration for a wide flat band, we exactly recover Eq.(\ref{eq:min}) and the mean-field result for large $N_\psi$.

\subsection {Random Phase Approximation} 
\begin{figure}
\centerline{\includegraphics[width=2.5in]{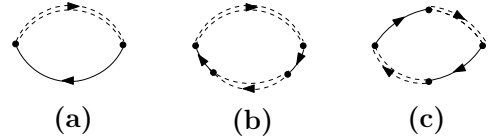}}
\caption{Various contributions to $\Sigma _b^{(2)}$. However, only diagram (a) contributes at order $1/N_\psi$. }
\label{fig:sec2}	
\end{figure}
The available expansion parameter in our model is $1/N_\psi$ and in order to justify the Gaussian saddlepoint (mean-field theory) of the previous section it is necessary to examine fluctuations systematically in a $1/N_\psi$ expansion. First we can examine several second order diagrammatic contributions to the Boson propagator,
\begin{equation}
{G_b}(i{\omega _n}) = {\left( {i{\omega _n} - \varepsilon _{\vec k}^b + {\Sigma _b}} \right)^{ - 1}}
\end{equation}
where $\Sigma _b$ is the exact self energy, which can be expanded perturbatively in the usual manner by including various virtual processes generated by Wick's theorem. Several of these self energy contributions are shown in Fig.\ref{fig:sec2} however only diagram (a) will contribute at order $1/N_\psi$. A proper renormalization of the perturbation series can be accomplished by summing a particular subset of the diagrams to infinite order as shown in Fig. \ref{fig:rpa}. This partial diagram summation is known as the Random phase approximation\cite{mattuck}, where in our case the bosonic interaction is effectively renormalized by a pseudo-dielectric response of the Fermion-molecule polarization bubble, which sums to:
\begin{equation}
\Delta {F_{RPA}} = \frac{1}{\beta }\sum\nolimits_n {\ln } \left[ { - i{\omega _n} + \varepsilon _{\vec k}^b + \Sigma _b^{(2)}(i{\omega _n})} \right]
\label{eq:rpacorr}
\end{equation}
and the irreducible Fermion-molecule bubble of Fig. \ref{fig:rpa} follows the usual frequency summation rules,
\begin{equation}
\begin{array}{l}
 \Sigma _b^{(2)}(i{\omega _n}) =\\ 
  = \frac{{{N_\psi }{{\left| V \right|}^2}}}{\beta }{\sum\limits_{k,\nu } {\frac{1}{{\left( {i{\omega _\nu } - i{\omega _n} - \varepsilon _k^f} \right)}}\left( {i{\omega _\nu } - \varepsilon _F^\psi  - \sum\limits_{k'} {\frac{{{{\left| {{{\tilde V}_{k'}}} \right|}^2}}}{{\left( {i{\omega _\nu } - \varepsilon _{k'}^f} \right)}}} } \right)} ^{ - 1}} \\ 
  = \frac{{{N_\psi }{{\left| V \right|}^2}}}{\beta }\sum\limits_{k,\nu } {\frac{1}{{\left( {i{\omega _\nu } - i{\omega _n} - \varepsilon _k^f} \right)}}\frac{1}{{\left( {i{\omega _\nu } - \varepsilon _F^\psi  - i\tilde \Delta {\mathop{\rm sgn}} ({\omega _\nu })} \right)}}}  \\ 
 \end{array}
 \label{eq:fermmol}
\end{equation}
which has been carefully evaluated by Read(1985) in the context of the the slave Boson representation of the infinite $U$ Anderson model\cite{read1985,hewsonbook, readnewns}, the details of which are shown explicitly in the appendix \ref{appa}. After taking the derivative of Eq.(\ref{eq:rpafinal}) with respect to $\tilde \Delta$, and adding the mean field contribution of Eq.(\ref{eq:mffinal}), the shift of the bosonic chemical potential to the leading order in $1/N_\psi$, at $T=0$  becomes:
\begin{figure}
\centerline{\includegraphics[width=3.0in]{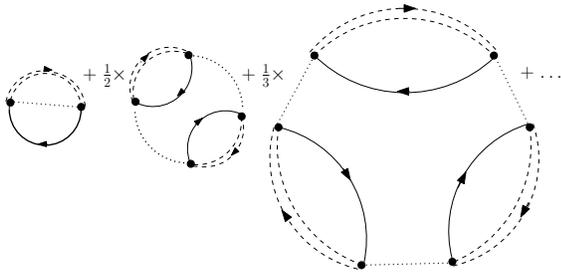}}
\caption{RPA-like corrections to the mean-field free energy }
\label{fig:rpa}	
\end{figure}

\begin{widetext}
\begin{equation}
\delta {\mu _b}(T = 0) = \frac{{N_\psi \Delta }}{\pi }\ln \left( {\frac{{\varepsilon _F^\psi }}{W}} \right) + \frac{{{{(N_\psi \Delta )}^2}}}{{N_\psi \pi }}\int_{ - W}^0 {d\epsilon } \left( {\frac{1}{{\epsilon  - \varepsilon _F^\psi }} + \frac{1}{{\varepsilon _F^\psi }}} \right)\left( {\epsilon  - \varepsilon _k^b - \frac{{N_\psi \Delta }}{\pi }\ln \left| {\frac{{\epsilon  - \varepsilon _F^\psi }}{{W }}} \right|} \right)^{-1}
\label{eq:mufinal}
\end{equation}
\end{widetext}
in which the integral of the RHS of Eq.(\ref{eq:mufinal}) renormalizes the chemical potential. Let us term this integral $I$ and focus on  its bandwidth $W$ dependence. $I$ can be conveniently re-expressed as,
\begin{equation}
I = \frac{{{\kappa ^2}\pi T_F^\psi }}{N}\int_0^{W/T_F^\psi } {\frac{x}{{1 + x}}\frac{{dx}}{{\left( {x + \varepsilon_k ^b/\varepsilon_F ^\psi +\kappa \ln (1 + x)} \right)}}} 
\end{equation}
where $\kappa  \equiv (N\Delta )/(\pi \varepsilon _F^\psi ) = (N\Delta )/(\pi {T_F ^\psi})$, for $k_B=1$. This expression can be integrated by parts, retaining the dominant leading order contribution, and dropping higher order logarithmic corrections,
\begin{equation}
I\sim \frac{{{\mu ^2}\pi }}{{{N_\psi }}}\frac{{W - \varepsilon _F^\psi \ln (1 + W/\varepsilon _F^\psi )}}{{W/\varepsilon _F^\psi  + \varepsilon _k^b/\varepsilon _F^\psi  + \kappa \ln (1 + W/\varepsilon _F^\psi )}}
\end{equation}

This yields the final expression,

\begin{equation}
\begin{aligned}
&\delta {\mu _b}(T = 0) \simeq \\
&\frac{{{N_\psi }\Delta }}{\pi }\ln \frac{{\varepsilon _F^\psi }}{W} + \frac{{{\mu ^2}\pi }}{{{N_\psi }}}\frac{{W - \varepsilon _F^\psi \ln (1 + W/\varepsilon _F^\psi )}}{{W/\varepsilon _F^\psi  - \varepsilon _k^b/\varepsilon _F^\psi  + \kappa \ln (1 + W/\varepsilon _F^\psi )}}
\end{aligned}
\label{eq:mulast}
\end{equation}

For the appropriate low temperature dependence one can again apply the Sommerfeld expansion as indicated earlier. Consequently, the convergence of the non-Gaussian fluctuations is verified and such corrections are sufficiently finite that mean field theory is stable at order $1/N_\psi$ and the quasiparticle description holds in the limit $N_\psi \to \infty$.
\section{Discussion: Chemical equilibrium and the Superfluid-Insulator phase transition} 
The presence of fermions will effect the superfluid-insulator transition of the Boson atoms at large $g_{bb}$. For $\mathcal{H}_{2p}$ type interactions this effect was explored by Refael and Demler, who found that the superfluid phase was suppressed by fermionic screening fluctuations\cite{refaelPRB2008}. On the other hand, Tewari \emph{et. al.} found a suppression of the Mott insulating lobes arising from a bosonic screening effect\cite{tewariPRB2009}. Consequently, the total effect is complex and the phase regions can be enhanced in either direction. 

For our particular system with finite $g_{3p}$, the renormalization of $\mu_b$ has important consequences in different physical phenomena which are experimentally measurable. In the preceding analysis we identified a characteristic crossover temperature $T_\psi$ that directly depends on the magnitude of $\varepsilon_F ^\psi$. Thus, by adjusting the molecular Fermi level $\varepsilon_F ^\psi$, one sets the crossover scale, given that our analysis is valid at temperatures $T<<T_\psi$. In accordance with the Kondo-like crossover, we expect $\mu_b$ to develop a local peak indicating that our approximation is breaking down, separating the low and high temperature regimes. This enhancement will affect the $\mu's$ and reaction rates of the other species, given that chemical equilibrium implies that the $\mu$'s of the components obey a rigid sum rule. Since the bosons contain a self interaction term, at very large repulsion the system is expected to undergo a transition into a Mott insulating state, and the mean-field phase diagram contains various insulating lobes separated by fixed values of the ratio $\mu_b$. The Mott insulating states are incompressible implying that
$\frac{{\partial {N_b}}}{{\partial {\mu _b}}} = 0$ and the phase boundary is governed by a zeros of some function $\phi (\mu_b /U)$. Thus, we predict qualitatively that by adjusting $\varepsilon_F ^\psi$ to smaller values, the phase boundaries of the lobes will be shifted downwards via the enhancement of $\mu_b$.

As mentioned earlier we have studied a model with tunable molecular formation. In the absence of resonant scattering there is a natural tendency for the atoms to form bound-states which relies upon the specific chemical affinities of the species. The resulting reaction rates are subject to classical chemical laws for the formation or disassociation based on the change in the Gibbs free energy. For a simple process like $\psi\mathbin{\lower.3ex\hbox{$\buildrel\textstyle\rightarrow\over{\smash{\leftarrow}\vphantom{_{\vbox to.5ex{\vss}}}}$}} f + b$ the conventional chemical thermodynamics in liquid solution using the law of mass action, relating the relative concentrations $\bar \rho_\nu$ and reaction rate $\mathcal{K}$ is given by the expression,
\begin{equation}
\mathcal{K} = \frac{{{{\bar \rho }_b}{{\bar \rho }_f}}}{{{{\bar \rho }_\psi }}} \sim {{\mathop{\rm e}\nolimits} ^{{\mathcal{E}_\psi}/{k_b}T}}
\end{equation}
which is a "thermally activated" process that limits $\mathcal{K}$ by the activation energy $\mathcal{E}_\psi$. In contrast, in ultra-cold atomic mixtures with tunable resonant interactions, the reactions rates are more complex, governed by Fermi's golden rule instead of the law of mass action, and the bare coupling $g_{3p}$ and subsequently $\mathcal{K}$ can be affected by vertex corrections and more subtle renormalization effects.

For an accurate physical interpretation, it is important to make note of certain differences between the Bose-Fermi mixture that we explored here and the slave Boson theory of the infinite $U$ Anderson model. Evidently, the final expression for the bosonic chemical potential corrections given by Eq.(\ref{eq:mulast}) does not contain any of the problematic infrared divergences of the Kondo problem which are related to the subtle restoration of gauge invariance. There are two primary reasons for this. First it must be made clear that we do not enforce any Lagrange multipliers onto the Hamiltonian (\ref{eq:mfham}), as the Bosons and Fermions are real physical particles that do not represent vacancies and occupied states. A Lagrange multiplier that reduces the span of the extended Hilbert space introduces an additional mean field parameter which renormalizes $\varepsilon_F ^\psi$.  Taking an additional derivative of the RPA correction of integral (\ref{eq:rpafinal}) with respect to the renormalized $\varepsilon_F ^\psi$ yields an infrared divergent integral. Furthermore, in our unconstrained system, the Bosons or the molecules do not arise from the decomposition of the Hubbard operators, ${\hat X_{m,0}} = \hat \psi _m^\dag \hat b$, ${\hat X_{0,m}} = {\hat \psi _m}{\hat b^\dag }$, ${\hat X_{0,0}} = {\hat b^\dag }\hat b$, ${\hat X_{m,m}} = \hat\psi _m^\dag {\hat\psi _m}$. Such a decomposition for a correlated electron systems retains a local gauge symmetry $\hat b\to \hat b e^{i\theta}$ ,$\hat b^\dag \to \hat b^\dag e^{i\theta}$ with a simultaneous change $\hat\psi\to \hat\psi e^{i\theta}$ ,$\hat\psi^\dag\to \hat\psi^\dag e^{i\theta}$. Because this symmetry is absent in our system, there should be no local gauge symmetry breaking issues.

One possible deficiency of our unconstrained mean-field theory was that we did not enforce a conservation of the volumes of the two Fermi surfaces. In particular, the conservation laws generated by Powell \emph{et al.} derives directly from Luttinger's theorem and a proper mean-field theory that reflects chemical equilibrium should enforce these constraints as Lagrange multipliers onto the Hamiltonian, thus generating a more complicated quasiparticle theory as observed in the Kondo problem, possibly generating an infrared divergence at order $1/N_\psi$.

By exploiting the single magnetic impurity framework by restricting the interactions to the hybridization of one degenerate molecular level with a wide flat atomic band we observe that the excitation spectrum of the Bose-Fermi mixture is qualitatively described by a Lorentzian broadened quasiparticle with width $\Delta$. The final result of Eq.(\ref{eq:mulast}) describes a $\mu_b$ controlled by the ratio  $\varepsilon_F ^\psi/W$, valid in the regime $\varepsilon_F ^\psi << W$, which is likely to impact the condensation behavior. Although we have specifically chosen an $N_\psi$-fold degenerate at the molecular Fermi surface, the analysis can be generalized to any energy or range of energies that are likely to hybridize with the atoms via $\mathcal{H}_{bf}^{3p}$. 

In conclusion, we have explored a wide variety of phenomena associated with the mean-field description of a tunable many-body chemical reaction that can be described in terms of a local Fermi liquid picture. These features can be probed directly by velocity and density distribution measurements via laser beam absorption\cite{blochrmp08}. Additionally, we have demonstrated a unique relationship between the condition for chemical equilibrium and the criteria for incompressibility observed in the Mott insulating phases of the superfluid-insulator transition for the Bosons.  Although we only considered a single molecular level it is conjectured that the certain qualitative features of the excitation spectrum of Hamiltonian (\ref{eq:hamilt1}), still admit a quasiparticle description, as the restoration of the momentum dependent self-energy corrections do not alter the Lorentzian broadened form of the mean-field Green's function given by Eq.(\ref{eq:mfgreen}) and subsequently the effective bosonic action should be renormalized in a similar fashion. Nevertheless, it is necessary to restore the explicit momentum dependence in the interactions to study the pairing bubbles of the self-energy that may detect screening/dielectric effects, and the novel possibility of nesting behavior and other Fermi surface instabilities that may indicate a breakdown of this quasiparticle picture.

\begin{acknowledgments}
I would like to give thanks to Yoichi Tanaka and Akira Furusaki for helpful conversations and assistance. This work was supported by the Institute for Physical and Chemical Research (RIKEN) FPR.
\end{acknowledgments}

\appendix
\section{Evaluation of $ \Sigma _b^{(2)}(i{\omega _n})$}

\label{appa}
Here we carry out the evaluation of the of the second order self-energy term $ \Sigma _b^{(2)}(i{\omega _n})$ as discussed in Refs.\cite{hewsonbook,read1985,readnewns}. Starting with Eq.(\ref{eq:fermmol}), where $\omega_\nu=(2\nu+1)\pi/\beta$, $\omega_n=2\pi n/\beta$ for integer values of $\nu$ and $n$, the $\omega_\nu$ integration can be expressed as a contour integral by applying the Matsubara summation procedure. We can approximate the atomic Fermion propagator as,
\begin{figure}
\centerline{\includegraphics[width=3.0in]{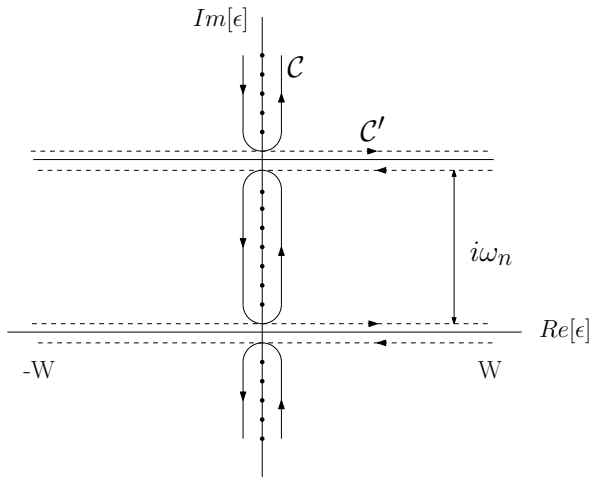}}
\caption{The contour $\mathcal{C}$ of Eq.(\ref{eq:contour}) is deformed to the contour $\mathcal{C'}$, allowing for a precise evaluation. }
\label{fig:bfcontour}	
\end{figure}
\begin{equation}
\sum\nolimits_k {{{\left( {i{\omega _n} + \varepsilon _k^f} \right)}^{ - 1}}}  \simeq  - i\pi \rho {\mathop{\rm sgn}} ({\omega _n})
\end{equation}
which for a constant density of states of width $2W$ yields the following contour integral, 
\begin{equation}
\begin{aligned}
 &\Sigma _b^{(2)}(i{\omega _n}) =  \\ 
  &= \frac{{N_\psi \Delta }}{2}\oint_{\mathcal{C}} {f(\epsilon )\frac{{\theta (W - \epsilon )\theta (W + \epsilon ){\mathop{\rm sgn}} ({\mathop{\rm Im}\nolimits} [\epsilon]  - {\omega _n})}}{{\epsilon  - \epsilon _F^\psi  + i\Delta {\mathop{\rm sgn}} ({\mathop{\rm Im}\nolimits} [\epsilon] )}}d\epsilon }  
 \end{aligned}
 \label{eq:contour}
\end{equation}
where the the contour $\mathcal{C}$ surrounds the poles on the imaginary axis a shown in Fig. \ref{fig:bfcontour}. The finite bandwidth introduces cuts that span $2W$ along the real axis, separated by the distance $i\omega_n$. The integral can be evaluated analytically at $T=0$ by distorting the contour $\mathcal{C} \to \mathcal{C'}$ which surrounds the cuts as shown in Fig. \ref{fig:bfcontour}.
This evaluates to,
\begin{equation}
\begin{aligned}
 &\Sigma _b^{(2)}(i{\omega _n}) = \frac{{N_\psi\Delta }}{\pi }\ln \left( {\frac{{i{\omega _n} - \varepsilon _F^\psi  + i\tilde \Delta {\mathop{\rm sgn}} ({\omega _n})}}{{ - \varepsilon _F^\psi  + i\tilde \Delta {\mathop{\rm sgn}} ({\omega _n})}}} \right) \\ 
  &+ \frac{{N_\psi \Delta }}{{2\pi }}\ln \left( {\frac{{{{(\varepsilon _F^\psi )}^2} + {{\tilde \Delta }^2}}}{{{W^2}}}} \right)  
 \end{aligned}
\end{equation}
Next to get the RPA free energy correction, the above expression is substituted into Eq.(\ref{eq:rpacorr}), and the Matsubara summation over the the Bose frequencies $\omega_n$ can be performed by extracting the imaginary part and deforming the contour to surround the negative real axis to yield,
\begin{widetext}
\begin{equation}
\Delta {F_{RPA}} =  - \frac{1}{\pi }\int_{ - W}^0 {{{\tan }^{ - 1}}\left[ {\frac{{(N_\psi \Delta /\pi )({{\tan }^{ - 1}}(\tilde \Delta /(\varepsilon _F^\psi  - \epsilon )) - {{\tan }^{ - 1}}(\tilde \Delta /\varepsilon _F^\psi ))}}{{\epsilon  - ({\varepsilon_k ^b } + (N_\psi \Delta /\pi )\ln \{ {{[{{(\varepsilon _F^\psi  - \epsilon )}^2} + {{\tilde \Delta }^2}]}^{1/2}}/W\} )}}} \right]d\epsilon } 
\end{equation}

To the leading order in $1/N_\psi$ this is simplified to yield,
\begin{equation}
\Delta F_{RPA}=\frac{-1}{\pi}\int_{ - W}^0 {d\varepsilon } \frac{\tilde \Delta }{\pi }\left( {\frac{1}{{\varepsilon  - \varepsilon _F^\psi }} + \frac{1}{{\varepsilon _F^\psi }}} \right)\left( {\varepsilon  - \varepsilon_k ^b - \frac{{N_\psi \Delta }}{\pi }\ln \left| {\frac{{\varepsilon  - \varepsilon _F^\psi }}{{W }}} \right|} \right)
\label{eq:rpafinal}
\end{equation}
\end{widetext}


\begin{thebibliography}{28}
\expandafter\ifx\csname natexlab\endcsname\relax\def\natexlab#1{#1}\fi
\expandafter\ifx\csname bibnamefont\endcsname\relax
  \def\bibnamefont#1{#1}\fi
\expandafter\ifx\csname bibfnamefont\endcsname\relax
  \def\bibfnamefont#1{#1}\fi
\expandafter\ifx\csname citenamefont\endcsname\relax
  \def\citenamefont#1{#1}\fi
\expandafter\ifx\csname url\endcsname\relax
  \def\url#1{\texttt{#1}}\fi
\expandafter\ifx\csname urlprefix\endcsname\relax\def\urlprefix{URL }\fi
\providecommand{\bibinfo}[2]{#2}
\providecommand{\eprint}[2][]{\url{#2}}

\bibitem[{\citenamefont{Bloch et~al.}(2008)\citenamefont{Bloch, Dalibard, and
  Zwerger}}]{blochrmp08}
\bibinfo{author}{\bibfnamefont{I.}~\bibnamefont{Bloch}},
  \bibinfo{author}{\bibfnamefont{J.}~\bibnamefont{Dalibard}}, \bibnamefont{and}
  \bibinfo{author}{\bibfnamefont{W.}~\bibnamefont{Zwerger}},
  \bibinfo{journal}{Rev. Mod. Phys.} \textbf{\bibinfo{volume}{80}},
  \bibinfo{eid}{885} (\bibinfo{year}{2008}).

\bibitem[{\citenamefont{Truscott et~al.}(2001)\citenamefont{Truscott, Strecker,
  McAlexander, Partridge, and Hulet}}]{truscott}
\bibinfo{author}{\bibfnamefont{A.~G.} \bibnamefont{Truscott}},
  \bibinfo{author}{\bibfnamefont{K.~E.} \bibnamefont{Strecker}},
  \bibinfo{author}{\bibfnamefont{W.~I.} \bibnamefont{McAlexander}},
  \bibinfo{author}{\bibfnamefont{G.~B.} \bibnamefont{Partridge}},
  \bibnamefont{and} \bibinfo{author}{\bibfnamefont{R.~G.} \bibnamefont{Hulet}},
  \bibinfo{journal}{Science} \textbf{\bibinfo{volume}{291}},
  \bibinfo{pages}{2570} (\bibinfo{year}{2001}).

\bibitem[{\citenamefont{Schreck et~al.}(2001)\citenamefont{Schreck, Khaykovich,
  Corwin, Ferrari, Bourdel, Cubizolles, and Salomon}}]{schreck}
\bibinfo{author}{\bibfnamefont{F.}~\bibnamefont{Schreck}},
  \bibinfo{author}{\bibfnamefont{L.}~\bibnamefont{Khaykovich}},
  \bibinfo{author}{\bibfnamefont{K.~L.} \bibnamefont{Corwin}},
  \bibinfo{author}{\bibfnamefont{G.}~\bibnamefont{Ferrari}},
  \bibinfo{author}{\bibfnamefont{T.}~\bibnamefont{Bourdel}},
  \bibinfo{author}{\bibfnamefont{J.}~\bibnamefont{Cubizolles}},
  \bibnamefont{and} \bibinfo{author}{\bibfnamefont{C.}~\bibnamefont{Salomon}},
  \bibinfo{journal}{Phys. Rev. Lett.} \textbf{\bibinfo{volume}{87}},
  \bibinfo{pages}{080403} (\bibinfo{year}{2001}).

\bibitem[{\citenamefont{Stan et~al.}(2004)\citenamefont{Stan, Zwierlein,
  Schunck, Raupach, and Ketterle}}]{stan}
\bibinfo{author}{\bibfnamefont{C.~A.} \bibnamefont{Stan}},
  \bibinfo{author}{\bibfnamefont{M.~W.} \bibnamefont{Zwierlein}},
  \bibinfo{author}{\bibfnamefont{C.~H.} \bibnamefont{Schunck}},
  \bibinfo{author}{\bibfnamefont{S.~M.~F.} \bibnamefont{Raupach}},
  \bibnamefont{and} \bibinfo{author}{\bibfnamefont{W.}~\bibnamefont{Ketterle}},
  \bibinfo{journal}{Phys. Rev. Lett.} \textbf{\bibinfo{volume}{93}},
  \bibinfo{pages}{143001} (\bibinfo{year}{2004}).

\bibitem[{\citenamefont{Inouye et~al.}(2004)\citenamefont{Inouye, Goldwin,
  Olsen, Ticknor, Bohn, and Jin}}]{inouye}
\bibinfo{author}{\bibfnamefont{S.}~\bibnamefont{Inouye}},
  \bibinfo{author}{\bibfnamefont{J.}~\bibnamefont{Goldwin}},
  \bibinfo{author}{\bibfnamefont{M.~L.} \bibnamefont{Olsen}},
  \bibinfo{author}{\bibfnamefont{C.}~\bibnamefont{Ticknor}},
  \bibinfo{author}{\bibfnamefont{J.~L.} \bibnamefont{Bohn}}, \bibnamefont{and}
  \bibinfo{author}{\bibfnamefont{D.~S.} \bibnamefont{Jin}},
  \bibinfo{journal}{Phys. Rev. Lett.} \textbf{\bibinfo{volume}{93}},
  \bibinfo{pages}{183201} (\bibinfo{year}{2004}).

\bibitem[{\citenamefont{DeMille and Hudson}(2008)}]{hudson}
\bibinfo{author}{\bibfnamefont{D.}~\bibnamefont{DeMille}} \bibnamefont{and}
  \bibinfo{author}{\bibfnamefont{E.~R.} \bibnamefont{Hudson}},
  \bibinfo{journal}{Nat. Phys.} \textbf{\bibinfo{volume}{4}},
  \bibinfo{pages}{911} (\bibinfo{year}{2008}).

\bibitem[{\citenamefont{Pilch et~al.}(2009)\citenamefont{Pilch, Lange,
  Prantner, Kerner, Ferlaino, N\"agerl, and Grimm}}]{pilchPRA2009}
\bibinfo{author}{\bibfnamefont{K.}~\bibnamefont{Pilch}},
  \bibinfo{author}{\bibfnamefont{A.~D.} \bibnamefont{Lange}},
  \bibinfo{author}{\bibfnamefont{A.}~\bibnamefont{Prantner}},
  \bibinfo{author}{\bibfnamefont{G.}~\bibnamefont{Kerner}},
  \bibinfo{author}{\bibfnamefont{F.}~\bibnamefont{Ferlaino}},
  \bibinfo{author}{\bibfnamefont{H.-C.} \bibnamefont{N\"agerl}},
  \bibnamefont{and} \bibinfo{author}{\bibfnamefont{R.}~\bibnamefont{Grimm}},
  \bibinfo{journal}{Phys. Rev. A} \textbf{\bibinfo{volume}{79}},
  \bibinfo{pages}{042718} (\bibinfo{year}{2009}).

\bibitem[{\citenamefont{Adhikari}(2004)}]{adhikariPRA04}
\bibinfo{author}{\bibfnamefont{S.~K.} \bibnamefont{Adhikari}},
  \bibinfo{journal}{Phys. Rev. A} \textbf{\bibinfo{volume}{70}},
  \bibinfo{pages}{043617} (\bibinfo{year}{2004}).

\bibitem[{\citenamefont{M\o{}lmer}(1998)}]{molmerPRL98}
\bibinfo{author}{\bibfnamefont{K.}~\bibnamefont{M\o{}lmer}},
  \bibinfo{journal}{Phys. Rev. Lett.} \textbf{\bibinfo{volume}{80}},
  \bibinfo{pages}{1804} (\bibinfo{year}{1998}).

\bibitem[{\citenamefont{{Amoruso} et~al.}(1998)\citenamefont{{Amoruso},
  {Minguzzi}, {Stringari}, {Tosi}, and {Vichi}}}]{amoruso98}
\bibinfo{author}{\bibfnamefont{M.}~\bibnamefont{{Amoruso}}},
  \bibinfo{author}{\bibfnamefont{A.}~\bibnamefont{{Minguzzi}}},
  \bibinfo{author}{\bibfnamefont{S.}~\bibnamefont{{Stringari}}},
  \bibinfo{author}{\bibfnamefont{M.~P.} \bibnamefont{{Tosi}}},
  \bibnamefont{and} \bibinfo{author}{\bibfnamefont{L.}~\bibnamefont{{Vichi}}},
  \bibinfo{journal}{Euro. Phys. J. D} \textbf{\bibinfo{volume}{4}},
  \bibinfo{pages}{261} (\bibinfo{year}{1998}).

\bibitem[{\citenamefont{B\"uchler and Blatter}(2004)}]{buchlerPRA04}
\bibinfo{author}{\bibfnamefont{H.~P.} \bibnamefont{B\"uchler}}
  \bibnamefont{and} \bibinfo{author}{\bibfnamefont{G.}~\bibnamefont{Blatter}},
  \bibinfo{journal}{Phys. Rev. A} \textbf{\bibinfo{volume}{69}},
  \bibinfo{pages}{063603} (\bibinfo{year}{2004}).

\bibitem[{\citenamefont{Das}(2003)}]{kkdasPRL03}
\bibinfo{author}{\bibfnamefont{K.~K.} \bibnamefont{Das}},
  \bibinfo{journal}{Phys. Rev. Lett.} \textbf{\bibinfo{volume}{90}},
  \bibinfo{pages}{170403} (\bibinfo{year}{2003}).

\bibitem[{\citenamefont{Cazalilla and Ho}(2003)}]{cazalillaPRL03}
\bibinfo{author}{\bibfnamefont{M.~A.} \bibnamefont{Cazalilla}}
  \bibnamefont{and} \bibinfo{author}{\bibfnamefont{A.~F.} \bibnamefont{Ho}},
  \bibinfo{journal}{Phys. Rev. Lett.} \textbf{\bibinfo{volume}{91}},
  \bibinfo{pages}{150403} (\bibinfo{year}{2003}).

\bibitem[{\citenamefont{Wang}(2006)}]{wangPRL2006}
\bibinfo{author}{\bibfnamefont{D.-W.} \bibnamefont{Wang}},
  \bibinfo{journal}{Phys. Rev. Lett.} \textbf{\bibinfo{volume}{96}},
  \bibinfo{pages}{140404} (\bibinfo{year}{2006}).

\bibitem[{\citenamefont{Lewenstein et~al.}(2004)\citenamefont{Lewenstein,
  Santos, Baranov, and Fehrmann}}]{lewensteinPRL2004}
\bibinfo{author}{\bibfnamefont{M.}~\bibnamefont{Lewenstein}},
  \bibinfo{author}{\bibfnamefont{L.}~\bibnamefont{Santos}},
  \bibinfo{author}{\bibfnamefont{M.~A.} \bibnamefont{Baranov}},
  \bibnamefont{and} \bibinfo{author}{\bibfnamefont{H.}~\bibnamefont{Fehrmann}},
  \bibinfo{journal}{Phys. Rev. Lett.} \textbf{\bibinfo{volume}{92}},
  \bibinfo{pages}{050401} (\bibinfo{year}{2004}).

\bibitem[{\citenamefont{Kokkelmans et~al.}(2002)\citenamefont{Kokkelmans,
  Milstein, Chiofalo, Walser, and Holland}}]{kokPRA2002}
\bibinfo{author}{\bibfnamefont{S.~J. J. M.~F.} \bibnamefont{Kokkelmans}},
  \bibinfo{author}{\bibfnamefont{J.~N.} \bibnamefont{Milstein}},
  \bibinfo{author}{\bibfnamefont{M.~L.} \bibnamefont{Chiofalo}},
  \bibinfo{author}{\bibfnamefont{R.}~\bibnamefont{Walser}}, \bibnamefont{and}
  \bibinfo{author}{\bibfnamefont{M.~J.} \bibnamefont{Holland}},
  \bibinfo{journal}{Phys. Rev. A} \textbf{\bibinfo{volume}{65}},
  \bibinfo{pages}{053617} (\bibinfo{year}{2002}).

\bibitem[{\citenamefont{Bruun and Pethick}(2004)}]{pethickPRL2004}
\bibinfo{author}{\bibfnamefont{G.~M.} \bibnamefont{Bruun}} \bibnamefont{and}
  \bibinfo{author}{\bibfnamefont{C.~J.} \bibnamefont{Pethick}},
  \bibinfo{journal}{Phys. Rev. Lett.} \textbf{\bibinfo{volume}{92}},
  \bibinfo{pages}{140404} (\bibinfo{year}{2004}).

\bibitem[{\citenamefont{Bortolotti et~al.}(2006)\citenamefont{Bortolotti,
  Avdeenkov, Ticknor, and Bohn}}]{bortJPB2006}
\bibinfo{author}{\bibfnamefont{D.~C.~E.} \bibnamefont{Bortolotti}},
  \bibinfo{author}{\bibfnamefont{A.~V.} \bibnamefont{Avdeenkov}},
  \bibinfo{author}{\bibfnamefont{C.}~\bibnamefont{Ticknor}}, \bibnamefont{and}
  \bibinfo{author}{\bibfnamefont{J.~L.} \bibnamefont{Bohn}},
  \bibinfo{journal}{J. Phys. B} \textbf{\bibinfo{volume}{39}},
  \bibinfo{pages}{189} (\bibinfo{year}{2006}).

\bibitem[{\citenamefont{Bortolotti et~al.}(2008)\citenamefont{Bortolotti,
  Avdeenkov, and Bohn}}]{bortPRA08}
\bibinfo{author}{\bibfnamefont{D.~C.~E.} \bibnamefont{Bortolotti}},
  \bibinfo{author}{\bibfnamefont{A.~V.} \bibnamefont{Avdeenkov}},
  \bibnamefont{and} \bibinfo{author}{\bibfnamefont{J.~L.} \bibnamefont{Bohn}},
  \bibinfo{journal}{Phys. Rev. A} \textbf{\bibinfo{volume}{78}},
  \bibinfo{eid}{063612} (\bibinfo{year}{2008}).

\bibitem[{\citenamefont{Powell et~al.}(2005)\citenamefont{Powell, Sachdev, and
  B\"uchler}}]{powellPRB05}
\bibinfo{author}{\bibfnamefont{S.}~\bibnamefont{Powell}},
  \bibinfo{author}{\bibfnamefont{S.}~\bibnamefont{Sachdev}}, \bibnamefont{and}
  \bibinfo{author}{\bibfnamefont{H.~P.} \bibnamefont{B\"uchler}},
  \bibinfo{journal}{Phys. Rev. B} \textbf{\bibinfo{volume}{72}},
  \bibinfo{pages}{024534} (\bibinfo{year}{2005}).

\bibitem[{\citenamefont{Fukuhara et~al.}(2007)\citenamefont{Fukuhara, Takasu,
  Kumakura, and Takahashi}}]{fukuhara}
\bibinfo{author}{\bibfnamefont{T.}~\bibnamefont{Fukuhara}},
  \bibinfo{author}{\bibfnamefont{Y.}~\bibnamefont{Takasu}},
  \bibinfo{author}{\bibfnamefont{M.}~\bibnamefont{Kumakura}}, \bibnamefont{and}
  \bibinfo{author}{\bibfnamefont{Y.}~\bibnamefont{Takahashi}},
  \bibinfo{journal}{Phys. Rev. Lett.} \textbf{\bibinfo{volume}{98}},
  \bibinfo{eid}{030401} (pages~\bibinfo{numpages}{4}) (\bibinfo{year}{2007}).

\bibitem[{\citenamefont{Abrikosov et~al.}(1975)\citenamefont{Abrikosov, Gorkov,
  and Dzyaloshinski}}]{abrikosov}
\bibinfo{author}{\bibfnamefont{A.~A.} \bibnamefont{Abrikosov}},
  \bibinfo{author}{\bibfnamefont{L.~P.} \bibnamefont{Gorkov}},
  \bibnamefont{and} \bibinfo{author}{\bibfnamefont{I.~E.}
  \bibnamefont{Dzyaloshinski}}, \emph{\bibinfo{title}{Methods of Quantum Field
  Theory in Statistical Physics}} (\bibinfo{publisher}{Dover Publications},
  \bibinfo{address}{New York}, \bibinfo{year}{1975}).

\bibitem[{\citenamefont{{Hewson}}(1997)}]{hewsonbook}
\bibinfo{author}{\bibfnamefont{A.~C.} \bibnamefont{{Hewson}}},
  \emph{\bibinfo{title}{The Kondo Problem to Heavy Fermions}}
  (\bibinfo{publisher}{Cambridge University Press},
  \bibinfo{address}{Cambridge}, \bibinfo{year}{1997}).

\bibitem[{\citenamefont{{Read}}(1985)}]{read1985}
\bibinfo{author}{\bibfnamefont{N.}~\bibnamefont{{Read}}}, \bibinfo{journal}{J.
  Phys. C} \textbf{\bibinfo{volume}{18}}, \bibinfo{pages}{2651}
  (\bibinfo{year}{1985}).

\bibitem[{\citenamefont{Read and Newns}(1983)}]{readnewns}
\bibinfo{author}{\bibfnamefont{N.}~\bibnamefont{Read}} \bibnamefont{and}
  \bibinfo{author}{\bibfnamefont{D.~M.} \bibnamefont{Newns}},
  \bibinfo{journal}{J. Phys. C} \textbf{\bibinfo{volume}{16}},
  \bibinfo{pages}{3273} (\bibinfo{year}{1983}).

\bibitem[{\citenamefont{{Mattuck}}(1977)}]{mattuck}
\bibinfo{author}{\bibfnamefont{R.~D.} \bibnamefont{{Mattuck}}},
  \emph{\bibinfo{title}{{A guide to Feynman diagrams in the many-body
  problem.}}} (\bibinfo{publisher}{Dover Publications}, \bibinfo{address}{New
  York}, \bibinfo{year}{1977}).

\bibitem[{\citenamefont{Refael and Demler}(2008)}]{refaelPRB2008}
\bibinfo{author}{\bibfnamefont{G.}~\bibnamefont{Refael}} \bibnamefont{and}
  \bibinfo{author}{\bibfnamefont{E.}~\bibnamefont{Demler}},
  \bibinfo{journal}{Phys. Rev. B} \textbf{\bibinfo{volume}{77}},
  \bibinfo{pages}{144511} (\bibinfo{year}{2008}).

\bibitem[{\citenamefont{Tewari et~al.}(2009)\citenamefont{Tewari, Lutchyn, and
  Das~Sarma}}]{tewariPRB2009}
\bibinfo{author}{\bibfnamefont{S.}~\bibnamefont{Tewari}},
  \bibinfo{author}{\bibfnamefont{R.~M.} \bibnamefont{Lutchyn}},
  \bibnamefont{and}
  \bibinfo{author}{\bibfnamefont{S.}~\bibnamefont{Das~Sarma}},
  \bibinfo{journal}{Phys. Rev. B} \textbf{\bibinfo{volume}{80}},
  \bibinfo{pages}{054511} (\bibinfo{year}{2009}).

\end{thebibliography}
\end{document}